\documentclass[english,aps,preprint]{revtex4}
\usepackage[LGR,T1]{fontenc}
\usepackage[latin9]{inputenc}
\usepackage{array}
\usepackage{float}
\usepackage{textcomp}
\usepackage{multirow}
\usepackage{stmaryrd}
\usepackage{graphicx}

\makeatletter

\DeclareRobustCommand{\greektext}{%
  \fontencoding{LGR}\selectfont\def\encodingdefault{LGR}}
\DeclareRobustCommand{\textgreek}[1]{\leavevmode{\greektext #1}}
\DeclareFontEncoding{LGR}{}{}
\DeclareTextSymbol{\~}{LGR}{126}
\providecommand{\tabularnewline}{\\}

\@ifundefined{textcolor}{}
{%
 \definecolor{BLACK}{gray}{0}
 \definecolor{WHITE}{gray}{1}
 \definecolor{RED}{rgb}{1,0,0}
 \definecolor{GREEN}{rgb}{0,1,0}
 \definecolor{BLUE}{rgb}{0,0,1}
 \definecolor{CYAN}{cmyk}{1,0,0,0}
 \definecolor{MAGENTA}{cmyk}{0,1,0,0}
 \definecolor{YELLOW}{cmyk}{0,0,1,0}
}

\makeatother

\usepackage{babel}
\begin{document}

\title{MAIN PARAMETERS OF LC$\varotimes$FCC BASED ELECTRON-PROTON COLLIDERS}

\author{Y. C. Acar}

\email{ycacar@etu.edu.tr}

\selectlanguage{english}%

\affiliation{TOBB University of Economics and Technology, Ankara, Turkey }

\author{U. Kaya}

\email{ukaya@etu.edu.tr}

\selectlanguage{english}%

\affiliation{TOBB University of Economics and Technology, Ankara, Turkey }

\affiliation{Department of Physics, Faculty of Sciences, Ankara University, Ankara,
Turkey}

\author{B. B. Oner}

\email{b.oner@etu.edu.tr}

\selectlanguage{english}%

\affiliation{TOBB University of Economics and Technology, Ankara, Turkey }

\author{S. Sultansoy}

\email{ssultansoy@etu.edu.tr}

\selectlanguage{english}%

\affiliation{TOBB University of Economics and Technology, Ankara, Turkey }

\affiliation{ANAS Institute of Physics, Baku, Azerbaijan}
\begin{abstract}
Multi-TeV center of mass energy ep colliders based on the Future Circular
Collider (FCC) and linear colliders (LC) are proposed and corresponding
luminosity values are estimated. Parameters of upgraded versions of
the FCC are determined to optimize luminosity of electron-proton collisions
keeping beam-beam effects in mind. It is shown that $L_{ep}\sim10^{32}\,cm^{-2}s^{-1}$
can be achieved with moderate upgrade of the FCC parameters.
\end{abstract}
\maketitle

\section{introduction}

Our knowledge on deep inside of matter has been essentially provided
by means of lepton-hadron collisions. For example, electron scattering
on atomic nuclei reveals structure of nucleons in Hofstadter experiment
\cite{key-1}. Moreover, quark parton model was originated from lepton
hadron-collisions \cite{key-2}. Extending the kinematic region by
two orders of magnitude both in high $Q^{2}$ and small x, HERA with
$\sqrt{s}=0.32$ TeV has shown its superiority compared to the fixed
target experiments and provided parton distribution functions (PDF)
for LHC experiments. Unfortunately, the region of sufficiently small
x ($<10^{-6}$) and high $Q^{2}$ ($\geq10\,GeV^{2}$), where saturation
of parton densities should manifest itself, has not been reached yet.
Hopefully, LHeC \cite{key-3} $\sqrt{s}=1.3$ TeV will give opportunity
to investigate this region. Construction of linear $e^{+}e^{-}$colliders
(or special linacs) and muon colliders tangential to the future circular
collider (FCC) as shown in Fig. 1 will give opportunity to achieve
multi-TeV center of mass energy in lepton-hadron collisions \cite{key-4-1}.

\begin{figure}[H]
\begin{centering}
\includegraphics[scale=0.5]{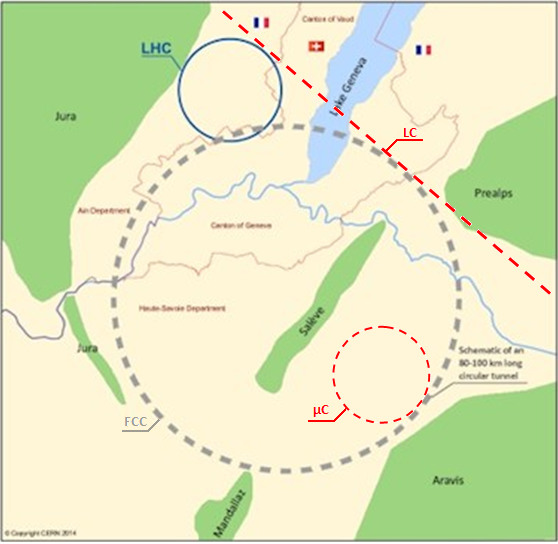}
\par\end{centering}

\centering{}\caption{Possible configuration for FCC, linear collider (LC) and muon collider
(\textmu C). }
\end{figure}

FCC is the future 100 TeV center-of-mass energy pp collider proposed
at CERN and supported by European Union within the Horizon 2020 Framework
Programme for Research and Innovation. Main parameters of the FCC
pp option \cite{key-4} are presented in Table I. The FCC also includes
an electron-positron collider option at the same tunnel (TLEP) \cite{key-5},
as well as several ep collider options \cite{key-6}.

\begin{table}[H]
\caption{Main parameters of proton beams in FCC.}

\centering{}%
\begin{tabular}{|c|c|}
\hline 
Beam Energy (TeV)  & 50\tabularnewline
\hline 
Peak Luminosity ($10^{34}\,cm^{-2}s^{-1}$) & 5.6\tabularnewline
\hline 
Particle per Bunch ($10^{10}$) & 10\tabularnewline
\hline 
 Norm. Transverse Emittance ($\mu m$) & 2.2\tabularnewline
\hline 
 \textgreek{b}{*} amplitude function at IP (m) & 1.1\tabularnewline
\hline 
IP beam size ($\mu m$) & 6.8\tabularnewline
\hline 
Bunches per Beam & 10600\tabularnewline
\hline 
Bunch Spacing (ns) & 25\tabularnewline
\hline 
 Bunch length (mm) & 80\tabularnewline
\hline 
Beam-beam tune shift  & $5.6\times10^{-3}$\tabularnewline
\hline 
\end{tabular}
\end{table}

Energy recovery linac (ERL) with $E_{e}=60\,GeV$ is chosen as the
main option for LHeC. Same ERL can also be used for FCC based ep collider
\cite{key-6}. Concerning e-ring in the FCC tunnel \cite{key-6} energy
of electrons is limited ($E_{e}<200\,GeV$) due to large synchrotron
radiation. Higher electron energies can be handled only by constructing
linear colliders tangential to the FCC. For the first time this approach
was proposed for UNK$\varotimes$VLEPP based ep colliders \cite{key-7}.
Then, construction of TESLA tangential to HERA was considered \cite{key-8}.
This line was followed by consideration of the LC$\varotimes$LHC
ep collider proposals (see reviews \cite{key-9,key-10} and references
therein). 

In this paper, we consider main parameters of the LC$\varotimes$FCC
based ep colliders. In numerical calculations, we use parameters of
ILC (International Linear Collider) \cite{key-11} and PWFA-LC (Plasma
Wake Field Accelerator - Linear Collider) \cite{key-12}. In Section
2 we estimate $L_{ep}$ taking into account beam-beam tune shift and
disruption effects. Finally, recommendations and conclusions are presented
in Section 3.

\section{LC$\varotimes$FCC based $ep$ Colliders}

General expression for luminosity of LC$\varotimes$FCC based ep colliders
is given by: 

$\,$

\[
L_{ep}=\frac{N_{e}N_{p}}{4\pi max[\sigma_{x_{p}},\sigma_{x_{e}}]max[\sigma_{y_{p}},\sigma_{y_{e}}]}min[f_{c_{p},}\,f_{c_{e}}]\quad(1)
\]

$\,$

where $N_{e}$ and $N_{p}$ are numbers of electrons and protons per
bunch, respectively; $\sigma_{x_{p}}$ ($\sigma_{x_{e}}$ ) and $\sigma_{y_{p}}$
($\sigma_{y_{e}}$ ) are the horizontal and vertical proton (electron)
beam sizes at IP; $f_{c_{e}}$ and $f_{c_{p}}$ are LC and FCC bunch
collision frequencies. $f_{c}$ is expressed by $f_{c}=N_{b}f_{rep}$,
where $N_{b}$ denotes number of bunches, $f_{rep}$ means revolution
frequency for FCC and pulse frequency for LC. In order to determine
collision frequency of ep collider, minimum value should be chosen
among electron and proton collision frequencies. Some of these parameters
can be rearranged in order to maximize $L_{ep}$ but one should note
that there are some main limitations that should be considered. One
of these limitations is electron beam power, however only parameters
of FCC proton beam is rearranged in this study and only nominal parameters
of linear colliders are considered. Therefore, there is no change
of electron beam power due to upgrades. Other limitations for linac-ring
type ep colliders are due to beam-beam effects. While beam-beam tune
shift affects proton beams, disruption has influence on electron beams.
Beam-beam tune shift for proton beams is given by:

$\,$

\[
\xi_{x_{p}}=\frac{N_{e}r_{p}\beta_{p}^{*}}{2\pi\gamma_{p}\sigma_{x_{e}}(\sigma_{x_{e}}+\sigma_{y_{e}})}\quad(2.1)
\]

$ $

\[
\xi_{y_{p}}=\frac{N_{e}r_{p}\beta_{p}^{*}}{2\pi\gamma_{p}\sigma_{y_{e}}(\sigma_{y_{e}}+\sigma_{x_{e}})}\quad(2.2)
\]

where $r_{p}$ is classical radius of proton, $\beta_{p}^{*}$ is
beta function of proton beam at interaction point (IP), $\gamma_{p}$
is the Lorentz factor of proton beam. $\sigma_{x_{e}}$ and $\sigma_{y_{e}}$
are horizontal and vertical sizes of electron beam at IP, respectively.
Disruption parameter for electron beam is given by: 

$\,$

\[
D_{x_{e}}=\frac{2\,N_{p}r_{e}\sigma_{z_{p}}}{\gamma_{e}\sigma_{x_{p}}(\sigma_{x_{p}}+\sigma_{y_{p}})}\quad(3.1)
\]

$\,$

\[
D_{y_{e}}=\frac{2\,N_{p}r_{e}\sigma_{z_{p}}}{\gamma_{e}\sigma_{y_{p}}(\sigma_{y_{p}}+\sigma_{x_{p}})}\quad(3.2)
\]

where $r_{e}$ is classical radius of electron, $\gamma_{e}$ is the
Lorentz factor of electron beam, $\sigma_{x_{p}}$ and $\sigma_{y_{p}}$
are horizontal and vertical proton beam sizes at IP, respectively.
$\sigma_{z_{p}}$ is bunch length of proton beam. 

Considering ILC$\varotimes$FCC and PWFA-LC$\varotimes$FCC options,
one should note that bunch spacing of electron accelerators are always
greater than FCC, while proton beam sizes are always greater than
the electron beam sizes at IP. Details and parameters of electron
beam accelerators are given in further subsections. In numerical calculations,
we use tranversely matched electron and proton beams at IP. Keeping
in mind roundness of FCC proton beam, Eqs (1)-(3) turn into;

$\,$

\[
L_{ep}=\frac{N_{e}N_{p}}{4\pi\sigma_{p}^{2}}f_{c_{e}}\quad(4)
\]

$\,$

\[
\xi_{p}=\frac{N_{e}r_{p}\beta_{p}^{*}}{4\pi\gamma_{p}\sigma_{p}^{2}}\quad(5)
\]

$\,$

\[
D_{e}=\frac{N_{p}r_{e}\sigma_{z_{p}}}{\gamma_{e}\sigma_{p}^{2}}\quad(6)
\]

$\,$

In order to increase luminosity of ep collisions moderate upgrade
of the FCC proton beam parameters have been used. Namely, number of
proton per bunch is increased 2.2 times, $\beta$-function of proton
beam at IP is arranged to be 11 times lower (0.1 m instead of 1.1
m) which corresponds to THERA and LHeC designs. Therefore, IP beam
size of proton beam, $\sigma_{p}$, is decreased 3.3 times according
to the relation $\sigma_{p}=\sqrt{\varepsilon_{p}^{N}\beta_{p}^{*}/\gamma_{p}}$.
Details of the parameter calculations for ep colliders are given in
further subsections.

\subsection{ILC$\varotimes$FCC}

Main parameters of ILC electron beam are given in Table II \cite{key-11}.
One can see from the table that bunch spacing of ILC is 554 ns which
is about 22 times greater than FCC bunch spacing of 25 ns. Therefore,
most of the proton bunches coming from FCC ring accelerator would
be dissipated unless parameters of FCC is rearranged. For FCC, the
parameter $N_{p}$ can be increased while number of bunches is decreased
regarding the dissipation. Transverse beam size of proton is much
greater than transverse beam size of electron for ILC$\otimes$FCC.
If beam sizes are matched, this leads $L_{ep}$ to decrease since
luminosity is inversely proportional to $\sigma_{p}^{2}$ as can be
seen from Eq. (4). To increase luminosity, upgraded value of $\beta_{p}^{*}$
parameter is set to be 0.1 m and therefore $\sigma_{p}$ to be 2.05
$\mu m$. Calculated values of $L_{ep}$, $D_{e}$ and $\xi_{p}$
parameters for ILC$\varotimes$FCC based ep colliders with both nominal
and upgraded FCC proton beam cases are given in Table 3. In addition
in Table 4, disruption parameter is fixed at the limit value of $D_{e}=25$
and corresponding $N_{p}$ and $L_{ep}$ values are given. 

\begin{table}[H]
\caption{Main parameters of electron beams in ILC.}

\centering{}%
\begin{tabular}{|c|c|c|}
\hline 
Beam Energy (TeV)  & $250$ & $500$\tabularnewline
\hline 
Peak Luminosity ($10^{34}\,cm^{-2}s^{-1}$) & $1.47$ & $4.90$\tabularnewline
\hline 
Particle per Bunch ($10^{10}$) & $2.00$ & $1.74$\tabularnewline
\hline 
 Norm. Horizontal Emittance ($\mu m$) & $10.0$ & $10.0$\tabularnewline
\hline 
Norm. Vertical Emittance (nm) & $35.0$ & $30.0$\tabularnewline
\hline 
Horizontal \textgreek{b}{*} amplitude function at IP (mm) & $11.0$ & $11.0$\tabularnewline
\hline 
Vertical \textgreek{b}{*} amplitude function at IP (mm) & $0.48$ & $0.23$\tabularnewline
\hline 
Horizontal IP beam size (nm) & $474$ & $335$\tabularnewline
\hline 
Vertical IP beam size (nm) & $5.90$ & $2.70$\tabularnewline
\hline 
Bunches per Beam & $1312$ & $2450$\tabularnewline
\hline 
Repetition Rate (Hz) & $5.00$ & $4.00$\tabularnewline
\hline 
Beam Power at IP (MW) & $10.5$ & $27.2$\tabularnewline
\hline 
Bunch Spacing (ns) & $554$ & $366$\tabularnewline
\hline 
Bunch length (mm) & $0.300$ & $0.225$\tabularnewline
\hline 
\end{tabular}
\end{table}

\begin{table}[H]
\caption{Main parameters of ILC$\varotimes$FCC based ep collider.}

\centering{}%
\begin{tabular}{|c|c|c|c|c|c|c|c|}
\hline 
 &  & \multicolumn{3}{c|}{Nominal FCC} & \multicolumn{3}{c|}{Upgraded FCC}\tabularnewline
\hline 
$E_{e}(GeV)$ & $\sqrt{s}(TeV)$ & $L_{ep}=10^{30}\,cm^{-2}s^{-1}$ & $D_{e}$ & $\xi_{p}$ & $L_{ep}=10^{30}\,cm^{-2}s^{-1}$ & $D_{e}$ & $\xi_{p}$\tabularnewline
\hline 
250 & 7.08 & 2.26 & 1.0 & $1.09\times10^{-3}$ & 55.0 & 24 & $1.09\times10^{-3}$\tabularnewline
\hline 
500 & 10.0 & 2.94 & 0.5 & $9.40\times10^{-4}$ & 70.0 & 12 & $9.40\times10^{-4}$\tabularnewline
\hline 
\end{tabular}
\end{table}

\begin{table}[H]
\caption{Main parameters of ILC$\varotimes$FCC based ep collider corresponding
to the disruption limit $D_{e}=25$. }

\centering{}%
\begin{tabular}{|c|c|c|c|c|}
\hline 
$E_{e}(GeV)$ & $\sqrt{s}(TeV)$ & $N_{p}(10^{11})$ & $L_{ep}=10^{30}cm^{-2}s^{-1}$ & $\xi_{p}$\tabularnewline
\hline 
250 & 7.08 & 2.3 & 57 & $1.09x10^{-3}$\tabularnewline
\hline 
500 & 10.0 & 4.6 & 149 & $9.40x10^{-4}$\tabularnewline
\hline 
\end{tabular}
\end{table}

\subsection{PWFA-LC$\varotimes$FCC}

Beam driven plasma wake field technology made a great progress for
linear accelerators recently. This method enables an electron beam
to obtain high gradients of energy even only propagating through small
distances compared to the radio frequency resonance based accelerators
\cite{key-12}. In other words, more compact linear accelerators can
be built utilizing PWFA to obtain a specified beam energy. In Table
V, main electron beam parameters of PWFA-LC accelerator are listed
\cite{key-12}. As in ILC$\varotimes$FCC case, transverse beam size
of proton is greater than all PWFA e-beam options. Same upgrade for
the proton beam is handled and final values of luminosity, disruption
and beam-beam parameters are given in Table VI for both nominal and
upgraded FCC proton beam cases. In Table VII, disruption parameter
is fixed at the limit value of $D_{e}=25$ and corresponding ep collider
parameters are given. 

\begin{table}[H]
\caption{Main parameters of electron beams in PWFA-LC.}

\centering{}%
\begin{tabular}{|c|c|c|c|c|c|}
\hline 
Beam Energy (TeV)  & 125 & 250 & 500 & 1500 & 5000\tabularnewline
\hline 
Peak Luminosity ($10^{34}\,cm^{-2}s^{-1}$) & 0.94 & 1.25 & 1.88 & 3.76 & 6.27\tabularnewline
\hline 
Particle per Bunch ($10^{10}$) & 1 & 1 & 1 & 1 & 1\tabularnewline
\hline 
Norm. Horizontal Emittance (m) & $1.00\times10^{-5}$ & $1.00\times10^{-5}$ & $1.00\times10^{-5}$ & $1.00\times10^{-5}$ & $1.00\times10^{-5}$\tabularnewline
\hline 
Norm. Vertical Emittance (m) & $3.50\times10^{-8}$ & $3.50\times10^{-8}$ & $3.50\times10^{-8}$ & $3.50\times10^{-8}$ & $3.50\times10^{-8}$\tabularnewline
\hline 
Horizontal \textgreek{b}{*} function at IP (m) & $11\times10^{-3}$ & $11\times10^{-3}$ & $11\times10^{-3}$ & $11\times10^{-3}$ & $11\times10^{-3}$\tabularnewline
\hline 
Veritcal \textgreek{b}{*} function at IP (m) & $9.9\times10^{-5}$ & $9.9\times10^{-5}$ & $9.9\times10^{-5}$ & $9.9\times10^{-5}$ & $9.9\times10^{-5}$\tabularnewline
\hline 
Horizontal IP beam size (m) & $6.71\times10^{-7}$ & $4.74\times10^{-7}$ & $3.36\times10^{-7}$ & $1.94\times10^{-7}$ & $1.06\times10^{-7}$\tabularnewline
\hline 
Veritfcal IP beam size (m) & $3.78\times10^{-9}$ & $2.67\times10^{-9}$ & $1.89\times10^{-9}$ & $1.09\times10^{-9}$ & $5.98\times10^{-10}$\tabularnewline
\hline 
Bunches per Beam & 1 & 1 & 1 & 1 & 1\tabularnewline
\hline 
Repetition Rate (Hz) & 30000 & 20000 & 15000 & 10000 & 5000\tabularnewline
\hline 
Beam Power ar IP (MW) & 6 & 8 & 12 & 24 & 40\tabularnewline
\hline 
Bunch Spacing (ns) & $3.33\times10^{4}$ & $5.00\times10^{4}$ & $6.67\times10^{4}$ & $1.00\times10^{5}$ & $2.00\times10^{5}$\tabularnewline
\hline 
Bunch length (m) & $2.00\times10^{-5}$ & $2.00\times10^{-5}$ & $2.00\times10^{-5}$ & $2.00\times10^{-5}$ & $2.00\times10^{-5}$\tabularnewline
\hline 
\end{tabular}
\end{table}

\begin{table}[H]
\caption{Main parameters of PWFA-LC$\varotimes$FCC based ep collider.}

\centering{}%
\begin{tabular}{|c|c|c|c|c|c|c|c|}
\hline 
 &  & \multicolumn{3}{c|}{Nominal FCC} & \multicolumn{3}{c|}{Upgraded FCC}\tabularnewline
\hline 
$E_{e}(GeV)$ & $\sqrt{s}(TeV)$ & $L_{ep}=10^{30}\,cm^{-2}s^{-1}$ & $D_{e}$ & $\xi_{p}$ & $L_{ep}=10^{30}\,cm^{-2}s^{-1}$ & $D_{e}$ & $\xi_{p}$\tabularnewline
\hline 
125 & 5.00 & 5.16 & 2.00 & $5.47\times10^{-4}$ & 124 & 48 & $5.47\times10^{-4}$\tabularnewline
\hline 
250 & 7.08 & 3.44 & 1.00 & $5.47\times10^{-4}$ & 82.6 & 24 & $5.47\times10^{-4}$\tabularnewline
\hline 
500 & 10.0 & 2.58 & 0.50 & $5.47\times10^{-4}$ & 61.9 & 12 & $5.47\times10^{-4}$\tabularnewline
\hline 
1500 & 17.3 & 1.72 & 0.17 & $5.47\times10^{-4}$ & 41.3 & 4.0 & $5.47\times10^{-4}$\tabularnewline
\hline 
5000 & 31.6 & 0.86 & 0.05 & $5.47\times10^{-4}$ & 20.8 & 1.2 & $5.47\times10^{-4}$\tabularnewline
\hline 
\end{tabular}
\end{table}

\begin{table}[H]
\caption{Main parameters of PWFA-LC$\varotimes$FCC based ep collider corresponding
to the disruption limit $D_{e}=25$. }

\centering{}%
\begin{tabular}{|c|c|c|c|c|c|c|}
\hline 
\multirow{2}{*}{$E_{e}(GeV)$} & \multirow{2}{*}{$\sqrt{s}(TeV)$} & \multirow{2}{*}{$N_{p}(10^{11})$} & \multirow{2}{*}{$L_{ep}=10^{30}cm^{-2}s^{-1}$} & \multirow{2}{*}{$\xi_{p}$} & \multicolumn{2}{c|}{IBS Growth Time (Horizontal) (h)}\tabularnewline
\cline{6-7} 
 &  &  &  &  & $L_{c}$=106.9 m & $L_{c}$=203.0 m\tabularnewline
\hline 
125 & 5.00 & 1.15 & 65.0 & $5.47\times10^{-4}$ & 721 & 149\tabularnewline
\hline 
250 & 7.08 & 2.30 & 86.0 & $5.47\times10^{-4}$ & 360 & 75.0\tabularnewline
\hline 
500 & 10.0 & 4.60 & 129 & $5.47\times10^{-4}$ & 180 & 37.0\tabularnewline
\hline 
1500 & 17.3 & 13.8 & 258 & $5.47\times10^{-4}$ & 60.0 & 12.0\tabularnewline
\hline 
5000 & 31.6 & 45.8 & 433 & $5.47\times10^{-4}$ & 18.0 & 3.90\tabularnewline
\hline 
\end{tabular}
\end{table}

As one can see from the third column of the Table VII number of protons
in bunches are huge in options corresponding to the highest energy
electron beams, therefore one may wonder about IBS growth times. For
this reason we estimate horizontal IBS growth times using Wei formula
\cite{key-13}. Obtained results are presented in the last two columns
of the Table VII. In numerical calculations we used FODO cell lengths
values $L_{c}$=106.9 m (same as LHC) and $L_{c}$=203.0 m considered
in \cite{key-14}. It is seen that IBS growth times are acceptable
even for $E_{e}=5000$ GeV case.

\section{Conclusions}

In this study it is shown that for ILC$\varotimes$FCC and PWFA-LC$\varotimes$FCC
based ep colliders luminosity values up to $L_{ep}\sim10^{32}\,cm^{-2}s^{-1}$
are achievable with moderate upgrade of the FCC proton beam. Even
with these luminosity values BSM search potential of ep colliders
essentially exceeds that of corresponding linear colliders and is
comparable with search potential of the FCC pp option for a lot of
BSM phenomena. In principle, ``dynamic focusing'' scheme \cite{key-15},
which was proposed for THERA, could provide $L_{ep}\sim10^{33}\,cm^{-2}s^{-1}$
for all ep collider options considered in this study. Concerning ILC$\varotimes$FCC
based ep colliders, a new scheme for energy recovery proposed for
higher-energy LHeC (see Section 7.1.5 in \cite{key-3}) may give an
opportunity to increase luminosity by an additional one or two orders,
resulting in $L_{ep}$ exceeding $10^{34}\,cm^{-2}s^{-1}$. Unfortunately,
this scheme can not be applied at PWFA-LC$\varotimes$FCC. Acceleration
of ion beams at the FCC \cite{key-14} will give opportunity to provide
multi-TeV center of mass energy in electron-nucleus collisions. In
addition, electron beam can be converted to high energy photon beam
using Compton back-scattering of laser photons (see \cite{key-16}and
references therein) which will give opportunity to construct LC$\varotimes$FCC
based $\gamma p$ and $\gamma A$ colliders (see \cite{key-17} and
references therein). 

In conclusion, construction of ILC and PWFA-LC tangential to the FCC
will essentially enlarge the physics search potential for both SM
and BSM phenomena. Therefore, systematic study of accelerator and
detector, as well as physics search potential, issues of LC$\varotimes$FCC
based lepton-hadron and photon-hadron colliders are essential to foreseen
the future of high energy physics.

\section{Acknowledgements}

This study is supported by TUBITAK under the grant No 114F337.

\end{document}